# DIRECT NON-BARYONIC DARK MATTER SEARCH
# - AN EXPERIMENTAL REVIEW -

Simon FIORUCCI

*DSM/DAPNIA, CEA Saclay, Bât. 141, Gif-sur-Yvette Cedex F-91191, France*
*sfiorucc@dapnia.cea.fr*

This review will present the latest advances in the search for non-baryonic dark matter from an experimental point of view, focusing more particularly on the direct detection approach. After a brief reminder of the main motivations for this search, we will expose the physical basis of WIMP detection, its advantages and limitations. The current techniques having achieved the most competitive results in terms of sensitivity will then be discussed. We will conclude with a rapid overview of the future of direct detection experiments, the techniques considered and their sensitivity goals.

1. Motivations

   1.1. Evidence for the existence of Dark Matter

The first proposal for the existence of a "dark" matter component in the universe goes back to 1933, with the observation by F. Zwicky[1] of velocity distributions in the Coma galactic cluster incompatible with the gravitational effect of visible light-emitting matter by two orders of magnitude. This observation was later extended to a large number of spiral galaxies by several teams, which all found a flat speed distribution at very large distances for visible objects in the galaxies arms, thus suggesting a spherical halo of non-visible matter surrounding the galaxies.

The first natural idea was to look for dark baryonic matter, such as unlit stars, Jupiter-like, cold gas clouds, or gravitational singularities like black holes. Results from a number of experiments – like EROS[2] or MACHO[3], however, have recently proved that such massive objects could only possibly amount for less than 20% of the total galactic halo mass.

Cosmological background observations have since then put the most drastic constraints upon the different cosmological parameters, notably WMAP in 2003[4]. Their conclusions are that within an apparently flat universe ($\Omega t = 1.00 +-0.02$), the total contribution of baryonic matter is $\Omega_b = 0.045 +-0.005$, while the total matter contribution is around 30%, thus leaving at least 85% of non-baryonic "dark" matter so far totally unaccounted for. Simulations indicate that most of this matter component must be non-dissipative, hence the denomination for this elusive matter component as Weakly Interacting Massive Particles, or WIMPs.

   1.2. WIMP candidates

One of the most natural WIMP candidates is provided by supersymmetric theories: If R-parity is conserved, the lightest supersymmetric particle is stable and leads to a relic density compatible with the missing mass component when the interaction cross-section is of the order of the weak interaction cross-section. Under the minimal supersymmetric standard model (MSSM) assumption, this LSP is often believed to be the neutralino, which is a linear combination of photino, zino and higgsino. Its predicted mass range is bound at low masses by LEP to 45 GeV[5], and at high masses to a few TeV, which gives it a non-negligible chance of being produced at LHC in the coming years. Rotation curves measurements lead to consider a WIMP halo of local density $\rho = 0.3$ GeV/cm$^3$, a maxwellian velocity distribution with $v_{rms} = 220$ km/s, and a relative velocity of the sun within the halo of 230 km/s.

Two ways of detecting such a particle have been considered. The indirect method is looking for products of WIMP annihilations such as neutrinos. Annihilations occurring preferentially in high density regions such as the center of the Sun, or to a lesser degree the center of the Earth, a strong directionality signature is expected. The SuperKamiokande[6], ANTARES[7] and AMANDA[8] experiments have obtained first results at the level of a few $10^{-6}$ pb for spin-independent interactions. The direct method, which will be further developed in this review, is looking directly for WIMP interactions within a target material.

2. The basis of direct detection

   2.1. Experimental signatures

The signal we expect from a WIMP contribution can bear several distinctive signatures. First comes the very nature of the observed interaction: Very preferentially, WIMPs, like neutrons, scatter elastically off a nucleus in the target material, while gammas and betas scatter off the electronic cloud. If one can devise a way to somehow differentiate in the data a nuclear recoil from an electronic recoil, a large part of the background will have thus been eliminated.

Secondly, for given WIMP characteristics (mass, cross-section) and for a given target material, a recoil energy spectrum is predicted, which can ideally be identified to the data, or at least allow to set an upper limit on the (mass, cross-section).

The WIMP halo is also supposed to be virialized in the galactic referential, with an average velocity $<V> \sim 220$ km/s. Due to the relative speed distributions between this halo and an observer on earth, accounting for the rotation of the Sun around the galactic center, of the Earth around the sun and of the Earth around its own axis, both an annual and diurnal modulation of the WIMP signal flux and directionality, respectively, are expected. The amplitude of the annual modulation can only amount for as high as a few percent of the total WIMP signal.

Lastly, a consistency between results acquired with different target materials, with different expected energy spectra, would certainly strengthen any eventual WIMP discovery.

### 2.2. Detection channels

For any material, however, the principles behind direct detection are always the same. It all comes down to measuring the energy deposited by a WIMP interaction – expected to be in the keV to tens of keV range. There are three ways of measuring this energy, which can eventually be combined. Each choice of a way of measure leads in turn to a choice of target material.

The fastest but less efficient way – in terms of collected energy – is to look for a light signal, produced by a scintillation process. This is used in liquid xenon or NaI crystals scintillators. The energy collection efficiency for this technique is typically of the order of a few percent.

With a few semi-conductor materials like Si, Ge, or CdTe e.g., one can also look for an ionization signal. Free charge carriers produced by the nuclear or electronic recoil can be collected onto electrodes using a drifting electrical field. The number of charges collected can then be related directly to the energy deposited. Depending on the material used and the operating conditions, the efficiency of this technique can be significantly higher than that of scintillation, of the order of a few tens of percents, though with a sensibly slower signal.

Lastly, a heat signal can be acquired, since almost all of the deposited energy eventually turns into heat one way or the other. This technique therefore usually ensures a higher energy collection efficiency, but to the cost of a slower signal, and usually tougher operating conditions – that is, cryogenics.

### 2.3. Challenges

Because of the very small interaction cross-section of WIMPs with ordinary matter, an extremely low event rate is expected – typically, less than a few events per kg of target material and per year for preferred SUSY models. Background suppression is therefore a vital issue. This is the reason why a number of experiments are burying themselves as deep as possible, to reduce the incident cosmic muon flux and cosmogenesis within their own detectors. Efforts are also made to ensure that the level of radio-purity of each component is kept to a minimum, and especially the parts closer to the detectors. Lastly, more or less complex setups can be used to effectively shield the detectors from natural radioactivity sources, such as will be presented later on in this paper.

All of this obviously becomes even truer when one considers the large mass scales, up to one ton, which are envisaged within the next ten years, while keeping a very low energy threshold of the order of the keV.

Furthermore, to effectively scan down to lower cross-sections, a further means of background discrimination appears to be more and more unavoidable. This can be achieved by using a technology which allows differentiating nuclear recoils events from electronic recoil events, thus eliminating most of the highly dominant gamma-ray background. One can also try to reduce the impact of "anomalous" events, for instance interactions occurring near the surfaces of crystal detectors and potentially able to give signals with different pulse shapes, or biased amplitudes. Such populations are already proving to be an effective limitation for a number of leading experiments.

3.  Current direct detection experiments

Taking into account all these parameters has led to a number of experiments set in various underground sites all over the world. In table 1 are reported the most notable or representative of them, which have already given results or are currently running. For ease of view, they are sorted by way of nuclear versus electronic recoil discrimination capabilities, into three categories: no discrimination capabilities, limited statistical discrimination capabilities and discrimination allowed on an event-by-event basis. We will now detail the principles and latest results for a few of them, trying to give the most comprehensive overview while avoiding technical lingo.

   3.1.  Non- or partially discriminative experiments

**DAMA**
The Italian experiment DAMA is set in the Gran Sasso laboratory, which gives it a protection of about 3500 meters-water-equivalent (mwe), thus reducing the incident muon flux by a factor $2.10^{-5}$, compared to sea level. The setup consists of 9 crystals of NaI scintillator, for a total mass of 100 kg. Details on its operating principle can be found in the paper by A. Incicchitti in this same volume.
DAMA is looking for an annual modulation of its total event rate, which would be a clear signature of the presence of WIMPs in the galactic halo, provided all other possibilities have been clearly ruled out. The expected amplitude of the modulation is about a few percent of the whole signal in the most favourable energy range. A limited statistical discrimination of nuclear and electronic recoils events is possible via pulse shape analysis, but this technique has not been used in the DAMA analysis since 1996.
Running since 1997, the data taking was completed in 2002 with a total exposure of over 107,731 kg.days, accumulated over 5 years. The published results[9] show a modulation confirmed at 6.3σ, using the 2-6 keV-electron-equivalent energy range, which has led the DAMA collaboration to claim model-independent evidence for the presence of WIMPs in the galactic halo (Figure1).

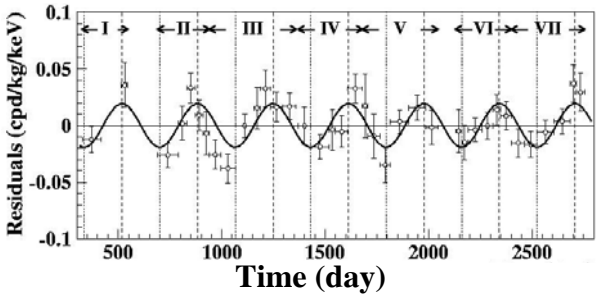

*Figure 1: annual modulation of the event rate as observed by the DAMA collaboration after more than 5 years of operation. x axis is time in days, y axis is residuals in cpd/kg/keV.*

Using standard halo parameters, this modulation translates to a WIMP candidate with:
$$M_\chi = 52^{+10}_{-8}\, GeV \quad \text{and} \quad \sigma_{\chi-n} = 7.2^{+0.4}_{-0.9} \cdot 10^{-6}\, pb.$$
The second phase of DAMA, LIBRA[10], is now running with 250 kg of NaI crystals.

**GENIUS-TF**
Also based in the Gran Sasso laboratory is the GENIUS-Test Facility experiment[11]. It uses 14 crystals of High Purity Germanium, 2.5 kg each, operated at liquid nitrogen temperature and placed behind heavy shielding layers of refined Ge, polystyrene, archaeological lead and borated polyethylene. The detectors are used as diodes, looking for an ionization signal only. GENIUS-TF aims to confirm or infirm the annual modulation observed by DAMA, which should be achievable within 5 years of operation, provided that the expected background of 0.01 event/kg/keV/day and the expected recoil energy threshold of 12 keV (3.5 keV electron equivalent) can be reached. This setup, however, incorporates no nuclear/electronic recoil discrimination capabilities.

The GENIUS-TF is really a prototype for the future double-beta decay and dark matter experiment GENIUS, which will operate 100kg of Germanium detectors.

**ZEPLIN**

The "ZonEd Proportional scintillation in Liquid Noble gas" experiment[12], as its acronym implies, uses liquid xenon as a scintillating material. Located in the Boulby mine in the UK, it comprises 5 kg of purified liquid xenon watched by 3 PMTs, and surrounded by a 1-ton Compton veto. The acquired light signal shows different time constants for a nuclear recoil event and an electronic recoil event. Both risetimes are also functions of the deposited energy, following a law which is ideally given by gamma and neutron calibrations.

The limited energy resolution (see below) implies that the nuclear recoil signal must be extracted by performing, for each energy bin, a Poisson analysis on the tail region of the electronic recoils risetime distribution. Assuming the distributions are really Gaussian, this gives an upper limit on the contribution of a WIMP signal.

With a data set of 230 kg.days, ZEPLIN claims to have achieved a sensitivity down to $10^{-6}$ pb for a WIMP of mass 60 GeV However, it should be noted that the data displays a resolution higher than 100% below 40 keV recoil, where most WIMP interactions would be observed. The overall background is also more than one order of magnitude higher than that of concurrent experiments. Reaching $10^{-6}$ pb would mean a background subtraction over 99.9 % in the relevant energy range, which in turns requires that the difference between risetime constants for both populations actually widens at low energies. The results from neutron calibrations, as well as results from other scintillating experiments, tend to imply that this has not been proven so far.

ZEPLIN has not yet published any definite results at this time.

### 3.2. Discriminative experiments

**CDMS**

Set in the Stanford shallow laboratory at 17 meters-water-equivalent depth, the phase I of the Cold Dark Matter Search experiment was offered relatively poor protection from cosmic rays, which it made up for with heavy shielding and an active muon veto efficient beyond 99 percent. CDMS uses 6 heat and ionization cryogenic germanium or silicium bolometers, with a technology which allows them to be sensitive to out-of-equilibrium phonons, and thus allows pulse shape analysis, as well as millimetric bi-dimensional impact position determination.

The data taking of CDMS-I is now completed, with a 28 kg.days data set already analyzed in 2003. It contains 20 nuclear recoil candidates, which could represent as many possible WIMP candidate interactions. However, it also displays at least 3 multiple interactions which can unambiguously be identified as neutrons, which means that CDMS-I is most probably limited by a neutron background. Two different exclusion limits, with and without neutron background subtraction[13], have thus been derived by the CDMS-I collaboration

The second phase of CDMS is now running in the Soudan mine, at 2090 mwe depth. After some cryogenics issues, 12 detectors are already taking data, and 18 more are in fabrication and to be installed soon. We are expecting their first results for summer 2004. Most recent preprints report an improvement in sensitivity by a factor eight compared to the 2003 result[23].

**CRESST**

The Cryogenic Rare Event Search with Superconducting Thermometers is using 300g CaWo4 Light and Heat bolometers. The heat signal is acquired by a small superconductive tungsten film thermometer attached to the crystal via a thin Si wafer, while the scintillation light is collected on a separate thin calorimeter. Each detector is surrounded by a light reflector to ensure higher light collection efficiency. The discrimination capabilities of this setup have been successfully tested beyond 99% above 15 keV of recoil energy, and a threshold of ~2 keV on the heat signal has been achieved. The most recent results with two detectors running already display a competitive sensitivity, yet clearly limited by the expected neutron background. The next phase of the experiment is now being installed, with eventually up to 33 detectors (10 kg) and additional protection against the neutron background.

**EDELWEISS**

The EDELWEISS experiment operates three 320g heat and ionization germanium bolometers at 17 mK. Set in the Modane underground laboratory, which offers an incident muon flux reduction by a factor $2.10^6$, the setup is further protected from radioactive backgrounds by several layers of shielding – 30cm paraffin, 20cm lead, 10cm copper.

The detectors themselves are germanium crystals, equipped with a 60 nm amorphous layer of germanium (GGA technology) or silicon (GSA), whose role is to limit anomalous surface events. 70 nm aluminium electrodes are sputtered on both faces of the crystal, and split into a centre electrode and a guard ring on one of the faces, for a better definition of the fiducial volume. A Neutron Transmutated Doped (NTD) germanium crystal is glued directly onto the germanium bulk and serves as a thermal sensor. Taking advantage of its two channels of measurement, EDELWEISS has announced discrimination capabilities between a nuclear recoil and an electronic recoil efficient beyond 99 % down to 15 keV of deposited energy.

The results of a first data taking were published in summer $2002^{14)}$, with a total exposure of 11.6 kg.days, resolutions of the order of the keV on each channel and zero event in the nuclear recoil band. This set the exclusion limit which was still at the time of this session the benchmark in terms of sensitivity to direct spin-independent WIMP interactions.

Nineteen more kg.days were added in 2003, but the data presented a handful of events in the nuclear recoil band. The derived result confirms but does not improve significantly the sensitivity reached in 2002. Since then, 21 more kg.days have been accumulated, using a new "phonon" trigger acquisition system, with improved resolutions. The data is still being analyzed. The experimental setup, meanwhile, is already being dismantled to make room for the next phase of the experiment.

3.3. Exclusion limits

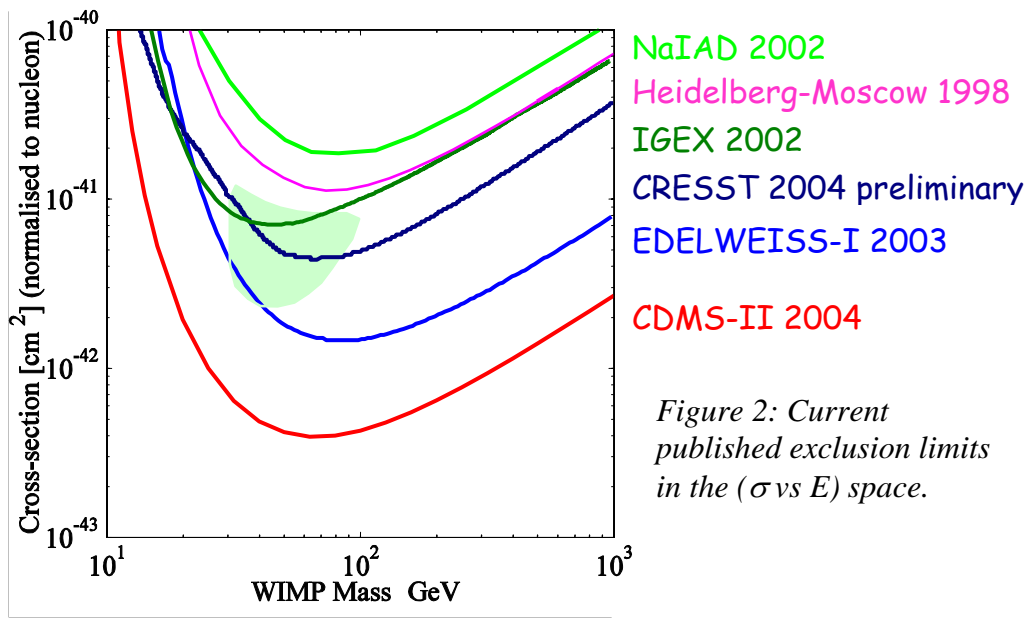

*Figure 2: Current published exclusion limits in the ($\sigma$ vs E) space.*

Figure 2 displays the current exclusion limits as they appear in the most recent published papers. Both the CDMS and EDELWEISS collaborations claim exclusion of the DAMA candidate – under the MSSM framework and assuming scalar WIMP interactions – at > 99.8 % confidence level, CDMS at low masses and EDELWEISS above 30 GeV. The new CDMS results further confirm this experimental contradiction. Recent publications have shown that this exclusion also resists to all realistic variations of the halo parameters. [21] [22]

4. Future experiments

Regardless of the technology, one of the key points for the next generation experiments will reside in a more precise understanding of the physics in the detectors, and a more systematic control of the background. This is why most future projects put an emphasis in their design on that particular point, while greatly increasing the detector mass. The second phases of CDMS, CRESST[15] and DAMA are already at least partially running. In table 2 are reported the most notable projects for direct dark matter detection experiments.
They can roughly be divided into three categories: cryogenic experiments, liquid xenon experiments, and more original designs. For each of these categories, we will now briefly present their main features through the example of a particular setup.

4.1. Cryogenic detectors

Displaying roughly the same kind of improvements as all other cryogenic experiments the second phase of EDELWEISS will see an increase in detector mass from 1 kg to eventually 35 kg of germanium bolometers, equivalent to 120 x 320g detectors. In this regard, a new cryostat has been designed and realized, and already successfully tested below 10 mK. In order to achieve better radioactive background levels, the radio-purity of all components has been thoroughly tested. An active muon veto will also help identify and reject the muon-induced fast neutron background.
In parallel, a new detector technology is being developed, based on thin $Nb_xSi_{1-x}$ electrodes able to collect at the same time an ionization signal and a fast out-of-equilibrium heat signal, thus allowing a better identification and rejection of anomalous surface events. [16]
Assembly in the laboratoire souterrain de Modane is starting in May 2004, and 20 to 30 fully operational detectors are expected to be taking data within a year. The ultimate goal of EDELWEISS-II is an improvement in sensitivity compared to phase-I of about 2 orders of magnitude.

4.2. Liquid Xenon experiments

A number of experiments using liquid xenon are being proposed, like the next phases of ZEPLIN[17], the Japanese XMASS[18] and the XENON[19] project. As the most ambitious of those, we will focus on the XENON project.
The XENON project is a dual phase xenon experiment using time projection chambers, contrary for example to ZEPLIN or XMASS which only use a liquid phase and can therefore only achieve a limited discrimination between electronic and nuclear recoils.
A WIMP interaction creates a primary scintillation light in the liquid phase, collected by a PMT array located in the gaseous phase. It also creates charge carriers which are drifted throughout the liquid phase, turned into proportional light by electroluminescence and collected by the same PMTs. A CsI photocathode deposited at the bottom of the chamber provides a "second primary" signal, and allows for some Z positioning.
A balloon-borne prototype with 8 cm drift has already been successfully operated, and a larger 10 kg prototype is underway, which will have to demonstrate the efficiency of the photocathode.
The final goal of the XENON experiment is the 1 ton scale, with a 16 keV recoil threshold and impressive background rejection capabilities. It aims at reaching $10^{-10}$ pb within 3 years.

### 4.3. Original designs

In order to try and circumvent some of the issues inherent to both cryogenic detectors and scintillators, alternative concepts are being looked into, trying for example to take advantage of different WIMP signatures. It is the case for example of the DRIFT[20] experiment, which uses time projection chambers to record the directionality of WIMP interactions. This should in theory make it sensitive to the diurnal modulation effect described in 2.1.

The setup consists in a projection chamber filled with a low pressure mixture of argon and $CS_2$, which serves as a charge drifting medium. The discrimination is based on the difference in recorded tracks ranges between different types of recoils, as well as pulse shape analysis. In effect, the nuclear recoil events are magnified by straight and short tracks, compared to electronic recoils at the same energy. The preliminary tests tend to show a very strong rejection of all gamma background down to a ten keV. After a number of technical difficulties, the DRIFT device has recently resumed testing.

### 5. Conclusions

All this leads to the predictive sensitivity diagram in figure 3.
Current experiments are nearing the $10^{-6}$ pb limit, and have already started to test the most optimistic super-symmetric models.
The next generation experiments, like EDELWEISS-II, CDMS-II, ZEPLIN-III, are aiming for an improvement of about 2 orders of magnitude, down to $10^{-8}$ pb which would allow them to test a much more significant part of the parameters space. This will come to the cost of an increased detector mass, further background rejection, lower energy thresholds and overall improved discrimination.

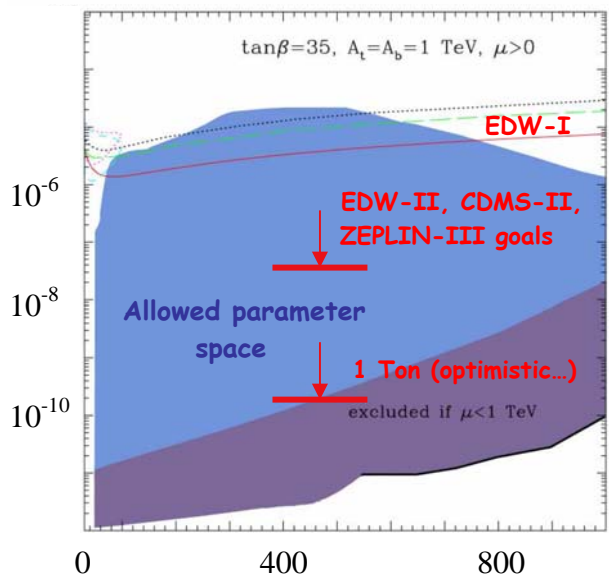

The 1-ton scale experiments are aiming yet two orders of magnitude lower, down to $10^{-10}$ pb, which would probe most of the allowed parameter space. Projections at this point remain hazardous, but it already appears clearly that both experimental issues and data analysis will prove very challenging. Combining results from different target materials, cross-checking with different signatures such as the directionality of interactions, may prove to be the only way to ever ascertain the existence of WIMPs with reasonable confidence.

*Figure 3: Summary predictive sensitivity plot. x axis is WIMP mass in GeV, y axis is WIMP cross-section in pb.*

*Table 1: Most notable current or finished direct dark matter search experiments*

| Discrim. | Name | Location | Technique | Material | Status |
|---|---|---|---|---|---|
| none | CUORICINO | Gran Sasso | Heat | 41 kg TeO2 | running |
| none | GENIUS-TF | Gran Sasso | Ionization | 42 kg Ge in N2 | running |
| none | HDMS | Gran Sasso | Ionization | 0.2 kg Ge diodes | stopped |
| none | IGEX | Canfranc | Ionization | 2 kg Ge Diodes | stopped |
| statitical | DAMA | Gran Sasso | Light | 100 kg NaI | stopped |
| statitical | LIBRA | Gran Sasso | Light | 250 kg NaI | running |
| statitical | NaIAD | Boulby mine | Light | 65 kg NaI | running |
| statitical | ZEPLIN-I | Boulby mine | Light | 4 kg Liquid Xe | running |
| Event-by-event | CDMS-I | Stanford | Heat + Ionization | 1 Kg Ge + Si | stopped |
| Event-by-event | CDMS-II | Soudan mine | Heat + Ionization | 2 to 7 kg Ge + Si | running |
| Event-by-event | CRESST-I | Gran Sasso | Heat + Light | 0.262 kg Al2O3 | stopped |
| Event-by-event | CRESST-II | Gran Sasso | Heat + Light | 0.6 to 9.9 kg CaWO$_4$ | running |
| Event-by-event | EDELWEISS-I | Modane | Heat + Ionization | 1 kg Ge | running |
| Event-by-event | ROSEBUD | Canfranc | Heat + Light | 1 kg BGO | running |
| Event-by-event | SIMPLE | Rustrel | Superheated droplets | Freon | stopped |

*Table 2: Future direct dark matter search experiments*

| Name | Location | Technique | Material | Status |
|---|---|---|---|---|
| ANAIS | Canfranc | Light | 10 to 100 kg NaI | R&D module running |
| CUORE | Gran Sasso | Heat | 760 kg TeO2 | Approved |
| DRIFT | Boulby mine | Ionization + Tracks | 1 m3 LPAr + CS2 | Testing |
| EDELWEISS-II | Modane | Heat + Ionization | 35 kg Ge | Upcoming early 2005 |
| XENON | Japan | Light + Ionization | 10 to 1000 kg LXe | R&D module running |
| XMASS | Japan | Light | 30 to 800 kg LXe | R&D module running |
| ZEPLIN-II | Gran Sasso | Light + Ionization | 30 kg LXe | Upcoming 2004 |
| ZEPLIN-III | Gran Sasso | Light + Ionization | 6 kg LXe | Upcoming 2005 |